\documentclass{phbauth}
\usepackage{graphicx}

\begin{document}

\begin{frontmatter}

\title{Interaction and Quantum Decoherence in Disordered Conductors}

\author[address1]{Andrei D. Zaikin\thanksref{thank1}},
\author[address2]{Dmitrii S. Golubev}

\address[address1]{Forschungszentrum Karlsruhe, Institut f\"ur
           Nanotechnologie,  D-76021 Karlsruhe, Germany}
\address[address2]{I.E.Tamm Department of Theoretical Physics, P.N.Lebedev
Physics Institute, Leninskii pr. 53, 117924 Moscow, Russia}
\thanks[thank1]{Corresponding author. E-mail: zaikin@int.fzk.de}

\begin{abstract}
We present a nonperturbative approach which allows to evaluate the
weak localization correction to the conductivity of disordered conductors
in the presence of interactions. The effect of the
electron-electron interaction on the magnetoconductance is described by the
function $A(t)\exp (-f(t))$. The dephasing time is determined only by 
$f(t)$, and this time remains finite down to $T=0$ due 
to the electron-electron interactions. In order to 
establish the relation between our nonperturbative analysis and the 
perturbative results the effect of interaction
on the pre-exponent $A(t)$ should be taken into account. The dephasing time 
cannot be unambiguously determined from a perturbative calculation.
\end{abstract}

\end{frontmatter}

\section{Introduction}

Recent experiments by Mohanty, Jariwala and Webb \cite{Webb} attracted a 
lot of attention to the fundamental 
role of interactions in disordered mesoscopic systems. These experiments
strongly indicate that the low temperature saturation of the 
electron decoherence time \cite{AAK,AAK1} $\tau_{\varphi}$ in 
disordered conductors has an intrinsic nature. Further analysis 
\cite{Mohanty} allowed to rule out various experimental artifacts
and several theoretical explanations based on extrinsic effects.

Can one expect the effect \cite{Webb} theoretically? 
In \cite{GZ} we developed a nonperturbative theoretical analysis of the above
phenomenon. We demonstrated that the low temperature saturation of the
decoherence rate $1/\tau_{\varphi}$ in disordered mesoscopic systems can have
an intrinsic nature: it can be caused by the electron-electron interaction. 
The results \cite{GZ} are in a good agreement with 
experimental findings \cite{Webb}.
We also argued \cite{GZ98} that this interaction-induced decoherence at low $T$
has the same physical nature as in the case of a quantum 
particle interacting with a bath of harmonic oscillators \cite{CL,LM,Weiss}.

In contrast, in \cite{AGA,AAG2} and in some other papers it was argued 
that interaction-induced dephasing of electrons at $T \to 0$ is not possible. 
Let us review the main arguments 
in favor of this conclusion \cite{AGA,AAG2}. 

One of such arguments is not necessarily related to electrons in a 
disordered metal. One can argue that a particle with energy $\sim T$ 
cannot excite harmonic oscillators with frequencies $\omega >T$ and, 
hence, the latter can only provide some renormalization effects.
Within this scenario electron scattering on such a static potential is purely 
elastic and cannot lead to dephasing. 

It is easy to observe 
that the above argument explicitly contradicts to the exact results
obtained e.g. within the Caldeira-Leggett model \cite{CL}. In this model 
even at $T=0$ the off-diagonal elements of the particle density matrix 
decay at 
a finite length set by interaction with an effective environment. 
This effect is due to {\it all} high frequency modes, i.e. the picture
is by 
no means ``static''. 
Our results \cite{GZ} demonstrate that also in disordered conductors 
the high frequency  ``quantum'' modes with $\omega >T$ {\it do} 
contribute to electron dephasing.

One can also argue that the electronic system can behave differently from a 
bosonic one \cite{CL} because of the Pauli principle which restricts the 
electron ability to exchange energy at low $T$. Again, this argument
contradicts to the well known results obtained for {\it fermionic} systems.
E.g. tunneling electrons exchange energy with
the effective environment (formed by other electrons in the leads) even at $T=0$ \cite{Naz,SZ}.
This so-called ``P(E)-theory'' \cite{Naz} yields measurable consequences and  
was verified in many experiments \cite{many}.
A close formal and physical similarity between the P(E)-theory and our
analysis \cite{GZ} is discussed in \cite{GZPE}.

A more formal argument is based on a perturbative calculation \cite{AAG2}.  
The authors \cite{AAG2} argued that their results explicitly
contradict to ours and, hence, the latter cannot be correct. A direct
comparison between the above results is not quite simple. 
The main reason for that is that our calculation \cite{GZ}
is essentially nonperturbative. It was performed with the exponential
accuracy which is sufficient to determine $\tau_{\varphi}$ 
and the weak localization correction to the conductance $\delta \sigma$
in the leading approximation for all values of the magnetic field.
Aleiner {\it et al.} \cite{AAG2} considered
the limit of strong magnetic fields $\tau_H \ll \tau_{\varphi}$ 
($\tau_H$ is the decoherence time due to the magnetic field) 
in which one can calculate the subleading
correction to $\delta\sigma $ perturbatively in the interaction. 
In this case the dephasing time $\tau_{\varphi}$ can be extracted from
a perturbative expression for $\delta \sigma (H)$ only if one
makes an additional assumption about the explicit form of the phase 
relaxation in time. No such assumption is needed within our analysis.
In addition to that, the
perturbative results \cite{AAG2} were presented only in the high temperature
limit $T\tau_H \gg 1$ which, being combined with $\tau_H \ll \tau_{\varphi}$,
is equivalent to a strong inequality 
$T\tau_{\varphi} >>> 1$. No such inequality was imposed in our calculation 
\cite{GZ}. 

In this paper we establish a direct relation between
perturbative and nonperturbative results in the problem of quantum dephasing
in disordered conductors. We will demonstrate that (i) the perturbation
theory in the interaction is insufficient for the problem
in question and (ii) in order to determine $\tau_{\varphi}$ it  
is necessary to perform a nonperturbative analysis with the exponential accuracy.
We analyze both the dephasing rate and the system magnetoconductance
for various temperatures and the magnetic fields.

\section{Insufficiency of the perturbation theory}

Originally the dephasing time $\tau_{\varphi}$ of electrons in disordered 
conductors was introduced and evaluated by Altshuler, Aronov and Khmelnitskii 
within a phenomenological nonperturbative procedure \cite{AAK}. Later Fukuyama and Abrahams
\cite{FA} studied the problem perturbatively in the interaction. Since
then the perturbative approach was frequently used in the analysis of 
quantum dephasing. We
are going to demonstrate that this perturbative strategy is insufficient and
does not allow to unambiguously determine the dephasing time even in 
those limits
where the weak localization correction to the conductance can be calculated
perturbatively in the interaction.    

In what follows we will restrict our analysis 
to quasi-1d disordered systems. Extension of our results to higher 
dimensions is straightforward \cite{GZ99}. In the 1d case the weak localization correction to the wire conductance
reads
\begin{equation}
\delta\sigma_1(H) = -\frac{e^2\sqrt{D}}{\pi^{3/2}}
\int\limits_0^{+\infty} \frac{dt}{\sqrt{t}} e^{-t/\tau_H}F(t/\tau_{\varphi}).
\label{1}
\end{equation}
Here the function $F(t/\tau_{\varphi})$ accounts for the 
electron-electron interaction ($F\equiv 1$ without interaction).
Provided in the long time limit the function $F$ decays faster than $1/\sqrt{t}$
the integral (\ref{1}) converges even for $1/\tau_H=0$ and we get
\begin{equation}
\delta\sigma_1 = -a\frac{e^2}{\pi}\sqrt{D\tau_\varphi},
\label{2}
\end{equation}
where $a \sim 1$ depends on the particular form of the function $F$. This result
is obviously nonperturbative in the interaction. E.g. if one formally defines
the ``interaction strength'' $\lambda$ in the effective Hamiltonian, in certain
limits one will
obtain $1/\tau_{\varphi} \propto \lambda$ and, hence, $\delta\sigma_1 \propto 
1/\sqrt{\lambda}$. For $\tau_H > \tau_{\varphi}$ any attempt to expand the 
conductance in powers of $\lambda$ is meaningless and may only yield 
divergences in all orders.

In order to avoid this problem the authors \cite{AAG2} suggested to consider the
limit of strong magnetic fields $\tau_H \ll \tau_{\varphi}$. In this case
the integral (\ref{1}) is cut at times $t \sim \tau_H$ much shorter than
$\tau_{\varphi}$ and the expansion of $\delta \sigma_1$ in powers of 
$\tau_H/\tau_{\varphi} \propto \lambda$ can be performed. Keeping only the 
leading terms of this expansion, from (\ref{1}) one readily finds
\begin{equation}
\delta\sigma_1 - \delta\sigma^{(0)}_1 \simeq
-\frac{e^2\sqrt{D\tau_H}}{2\pi}\frac{\tau_H}{\tau_{\varphi}}F'(0),
\label{2a}
\end{equation}
where $\delta\sigma^{(0)}_1 = -(e^2/\pi)\sqrt{D\tau_H}$. We observe that a
perturbative expansion in the interaction allows us to determine
the combination $F'(0)/\tau_{\varphi}$ but not $\tau_{\varphi}$ itself.
The full function $F(t/\tau_{\varphi})$ (and thus $F'(0)$) will remain
unknown. Although in the limit $\tau_H \ll \tau_{\varphi}$ the value
$\delta \sigma_1(H)$ can be calculated perturbatively in the interaction,
this would yield no information about the dephasing time $\tau_{\varphi}$.
Such information can be extracted from the perturbation theory only if one {\it assumes}
some particular form of the function $F(t/\tau_{\varphi})$. But this form should
be {\it found} rather than assumed. This task can be accomplished only if one 
goes beyond the perturbation theory.

In order to illustrate this conclusion let us consider several different functions
$F(t/\tau_{\varphi})$. Perhaps the most frequent choice of this function is
\begin{equation}
F_1(t/\tau_{\varphi}) = \exp (-t/\tau_{\varphi}).
\label{1a}
\end{equation}
This {\it assumption} was also adopted in \cite{AAG2}. Performing a perturbative 
calculation in the high temperature limit $T\tau_H \gg 1$ the authors \cite{AAG2} 
arrived at the result $\delta\sigma_1 - \delta\sigma^{(0)}_1 \propto T\tau_H^2$
which, being combined with (\ref{2a}) and (\ref{1a}), yields the dephasing rate 
\cite{AAG2} $1/\tau_{\varphi 1} \propto T\sqrt{\tau_H}$.

Another possible choice of the function $F$ can be
\begin{equation}
F_2(t/\tau_{\varphi}) = \exp (-(t/\tau_{\varphi})^{3/2}).
\label{1b}
\end{equation}
For $\tau_H >\tau_{\varphi}$ this function -- as well as (\ref{1a}) -- yields eq. (\ref{2}).
However, in the limit $\tau_H \ll\tau_{\varphi}$ from (\ref{1}) and (\ref{1b}) one finds
\begin{equation}
\delta\sigma_1- \delta\sigma^{(0)}_1
\simeq
\frac{e^2\sqrt{D\tau_H}}{\pi^{3/2}}\left(\frac{\tau_H}{\tau_{\varphi}}\right)^{3/2}.
\label{2b}
\end{equation}
Clearly, this result is {\it incompatible} with eq. (\ref{2a}) because $F'(0)$ 
should not depend on $\tau_H$ and $\tau_{\varphi}$. At the same time from (\ref{1b}) one has
$F'(0)=0$, and again the contradiction between (\ref{2a}) and (\ref{2b}) is obvious.
Also, combining (\ref{2b}) with the perturbative results \cite{AAG2} one arrives at
$1/\tau_{\varphi 2} \propto T^{2/3}$ in agreement with \cite{AAK,AAK1} but in a
clear disagreement with $1/\tau_{\varphi 1}$ found in \cite{AAG2}.

Finally, let us choose the trial function $F$ in the following form:
\begin{equation}
F_3(t/\tau_{\varphi}) = \frac{e^{-t/\tau_\varphi}\sqrt{bt}}
{\sqrt{\tau_\varphi(1-e^{-bt/\tau_\varphi})}},
\label{1c}
\end{equation}
where $b \sim 1$. For $\tau_H > \tau_{\varphi}$ the result (\ref{2}) is recovered
again, while in the limit $\tau_H \ll \tau_{\varphi}$ one arrives at eq. (\ref{2a})
with $F'(0)=b/4-1$. Combining this result with those of \cite{AAG2}
one finds $\tau_{\varphi 3} \propto (4-b)/(T\tau_H^{1/2})$. For 
$b<4$, $b=4$ and $b>4$ this result would imply respectively positive,
infinite and even negative dephasing rates, which is an obvious nonsense.

The above examples clearly demonstrate that a perturbative
procedure is {\it principally} insufficient for our problem 
because it leads to completely ambiguous results for
$\tau_{\varphi}$. At the same time the nonperturbative eq. (\ref{1})
yields practically indistinguishable magnetoconductance curves (see Fig. 1 of \cite{GZ99})
and {\it the same} dephasing time $\tau_{\varphi}$ (up to a prefactor $a \sim 1$ which
can be absorbed in $\tau_{\varphi}$ anyway) for all the trial functions $F_1$, $F_2$ 
and $F_3$.

In order to obtain correct information about $\tau_{\varphi}$ one should 
determine the function $F$ at times $t \sim \tau_{\varphi}$. This can be done only 
by means of a nonperturbative calculation simply
because for $t \sim \tau_{\varphi}$ there exists no small parameter in the problem.
Choosing the limit $\tau_H \ll \tau_{\varphi}$ enables one to find 
$\delta \sigma_1 (H)$ but {\it not} $\tau_{\varphi}$. We can also add that
the phenomenological procedure \cite{AAK} allows to non-perturbatively treat only the contribution
of ``classical'' modes $\omega < T$, whereas 
the effect of ``quantum'' modes
$\omega >T$ is accounted for by means of the nonperturbative analysis 
\cite{GZ}. This analysis will be extended further in the next section.

\section{Nonperturbative results}

Let us express the function $F$ (\ref{1}) in the form
\begin{equation}
F(t)=A(t)\exp (-f(t))
\label{F}
\end{equation}
Without interaction one has $A(t) \equiv 1$ and $f(t) \equiv 0$.
For a complete description of the effect of
interaction on the weak localization correction (\ref{1}) it is
in general necessary to evaluate both functions $f(t)$ and $A(t)$.
An important observation is, however, that information about the
effect of interaction on $A(t)$ is not needed to correctly evaluate
the dephasing time $\tau_{\varphi}$, it is sufficient to find only the function
$f(t)$ which describes the decay of correlations in time and provides
an effective cutoff for the integral (\ref{1}) at $t \sim \tau_{\varphi}$.
The role of the pre-exponent $A(t)$ is merely to establish an exact numerical
prefactor. Since $\tau_{\varphi}$ is defined up to a numerical prefactor
of order one anyway, it is clear that only the function $f(t)$ 
-- and not $A(t)$ -- is really important.

\subsection{Exponent}

The function $f(t)$ can be straightforwardly evaluated by means of
the path integral formalism. This procedure amounts to calculating the
path integral for the kernel of the evolution operator over the particle
coordinate $r$ and the momentum $p$ \cite{GZ}
\begin{equation}
J \sim \int D r \int D p \, \exp (iS_0-iS_0'-iS_R-S_I)
\label{10}
\end{equation}
within the saddle point approximation on pairs of time reversed paths and to averaging
over diffusive trajectories. Here $S_0$ and $S_0'$ represent the electron action
on the two parts of the Keldysh contour, while $iS_R+S_I$ accounts for the interaction.
The general expressions for $S_R$ and $S_I$ were derived in \cite{GZ}. Within
RPA these expressions contain the full information about the effect of interaction
in {\it all} orders of the perturbation theory.

The saddle point approximation procedure was
also described in details in \cite{GZ}. One can demonstrate 
that the contribution of the real part $S_R$ of the action  vanishes on
any pair of diffusive time reversed paths. Note that such cancellation is
a generic property of a wide class of influence functionals describing
dissipative environments. E.g. the same cancellation is observed in the Caldeira-Leggett
model \cite{CL} and in some other models \cite{Weiss}.
Hence, the function $f(t)$ in the exponent (\ref{F}) is determined only by the
imaginary part of the action $S_I$ \cite{GZ}. For the sake of generality here we
will present the result for all dimensions. Evaluating $S_I$ on pairs
of time reversed saddle point paths we find \cite{GZ99}
$$
f(t)=\frac{4e^2D^{1-d/2}}{\sigma_d (2\pi)^d}
\int\frac{d^dx}{1+x^4} 
\int\frac{d\omega\; d\omega'(1-\cos\omega t)}{(2\pi)^2}
$$
$$
\times \left[\frac{|\omega'|^{d/2-2}(\omega-\omega')\coth\frac{\omega-\omega'}{2T}
}{\omega^2}+
\right.
$$
\begin{eqnarray}
\left.
+
\frac{|\omega'|^{d/2-2}\omega\coth\frac{\omega}{2T}+
|\omega|^{d/2-2}\omega'\coth\frac{\omega'}{2T}}{\omega^{\prime 2}-\omega^2}
\right].
\label{fott}
\end{eqnarray}
Note that in 1d and 2d cases the integral of the first term over $\omega'$
diverges at $\omega'\to 0$. However, this divergence
is exactly canceled by
the second term and the whole integral is finite in any dimension. 

Evaluating the result (\ref{fott}) for a 1d case  
in the quantum regime $\pi Tt\ll 1$ we obtain
\begin{equation}
f(t)\simeq \frac{e^2}{\pi\sigma_1}\sqrt{\frac{2D}{\tau_e}}t+
\frac{2e^2}{\pi\sigma_1}\sqrt{\frac{Dt}{\pi}}\left(
\ln\frac{2\pi t}{\tau_e}-6\right),
\label{fquantum}
\end{equation}
where $\tau_e=v_F/l$ is the elastic mean free time. 
Note, that apart from the leading linear in time term
there exists a smaller term $\propto \sqrt{t}\ln(t/\tau_e)$, which also grows
in time.
In the opposite thermal limit $\pi Tt\gg 1$ eq. (\ref{fott}) yields 
\begin{eqnarray}
f(t) \simeq \frac{e^2}{\pi\sigma_1}\sqrt{\frac{2D}{\tau_e}}\left[
t+\frac{2\sqrt{2\pi\tau_e}}{3}Tt^{3/2}
\right.
\nonumber\\
\left.
+\frac{\pi^{1/2}\zeta(1/2)}
{\sqrt{2}}t\sqrt{T\tau_e}
\right],
\label{fthermal}
\end{eqnarray}
where $\zeta (x)$ is the dzeta-function. We observe that in 
both cases (\ref{fquantum}) and (\ref{fthermal}) there
exists a linear in time temperature independent contribution to $f(t)$
which determines the dephasing time $\tau_{\varphi}$ at low temperatures \cite{GZ,GZ98}.
Beside that at $Tt \gg 1$ there exists another term $\propto Tt^{3/2}$
which yields dominating contribution to $\tau_{\varphi}$ at high temperatures
$T > T_q \sim 1\sqrt{\tau_{\varphi}\tau_e}$,
where the result \cite{AAK} $\tau_{\varphi} \propto T^{-2/3}$ is recovered.

\subsection{Pre-exponent}

A rigorous calculation of the pre-exponential function 
$\tilde{A}(t)=A(t)/\sqrt{t}$ in (\ref{1}) in the presence of interactions 
is beyond the frames of the present paper. Fortunately the precise 
form of $\tilde{A}(t)$ for all times 
is not interesting for us here. Of importance is to qualitatively 
understand how the function $\tilde{A}(t)$ is modified in the presence 
of the electron-electron interaction. 

It is well known \cite{AAK1,CS} that without interactions the function 
$\tilde{A}(t)$
is related to the return probability of diffusive trajectories to the same
point after the time $t$. In the presence of dissipation (described by the term
$S_R$ in the effective action) the particle energy decreases and its diffusion
slows down. This implies that at any given time $t$ the function 
$\tilde{A}(t)$ should exceed $1/\sqrt{t}$. On the other hand, at least if 
the interaction is sufficiently weak, diffusion will take place at 
all times and, hence, 
$\tilde{A}(t)$ will always decay as $1/\sqrt{t}$ or slower.

The latter -- intuitively obvious -- property of the pre-exponent 
was confirmed by our analysis \cite{GZ99}. The function $\tilde{A}(t)$ indeed
decays at all times and no compensation of the exponential 
decay $\propto \exp (-f(t))$ by the pre-exponential
function $\tilde{A}(t)$ can occur in the long time limit. Another important 
property \cite{GZ99} is that in the interesting limit of low
temperatures the effect of interaction on the pre-exponent becomes important
at $t \sim \tau_{\varphi}$, i.e. on the same time scale as for the function 
$f(t)$ in the exponent. These two properties allow to completely ignore
the effect of interaction on the pre-exponent \cite{GZ}.

In the short time limit $t \ll \tau_{\varphi}$ the correction to the 
pre-exponent due to interaction
is small and one can proceed perturbatively in the interaction. 
At $T \to 0$ in the leading approximation one finds
\cite{GZ99}
\begin{equation}
A(t)\simeq 1+ \frac{e^2}{\pi\sigma_1}\sqrt{\frac{2D}{\tau_e}}t.
\label{Acorr}
\end{equation}
This equation is important for deriving the perturbative
expressions for the magnetoconductance in the limit $\tau_H \ll \tau_{\varphi}$.
The corresponding results are presented below.

\section{Perturbation theory for the conductance} 

Let us evaluate the weak localization correction (\ref{1}) in the limit of strong
magnetic fields $\tau_H \ll \tau_{\varphi}$. Performing the short time 
expansion of both $\exp (-f(t))$ and $A(t)$ to the first order and 
combining (\ref{1}), (\ref{F}) with (\ref{fquantum}), 
(\ref{Acorr}), in the limit $T\tau_H \ll 1$ we find
\begin{equation}
\delta\sigma_1 - \delta\sigma^{(0)}_1 \simeq \frac{e^2}{\pi}\frac{e^2}{\sigma_1}
\frac{2D\tau_H}{\pi^2}\left(\ln \left(\frac{\tau_H}{\tau_e}\right)-3.74\right).
\label{GZ156}
\end{equation} 
It is easy to observe that this result does not contain the zero temperature
dephasing time at all. The linear in time terms in the expressions for $f(t)$
(\ref{fquantum}) and $A(t)$ (\ref{Acorr}) cancel each other exactly in the
first order. The same cancellation occurs in the limit $T\tau_H \gg 1$. Again
combining (\ref{1}), (\ref{F}) with (\ref{fthermal}), (\ref{Acorr}) 
and expanding
$\exp (-f(t))$ to the first order in $f$, at $T\tau_H \gg 1$ we obtain
\begin{equation}
\delta\sigma_1 -\delta \sigma_1^{(0)}\simeq \frac{e^2}{\pi}\frac{e^2}{\sigma_1}DT\tau_H^2
\left[\frac{4}{3\pi}+\frac{\zeta(1/2)}{2\sqrt{2\pi T\tau_H}} \right].
\label{GZ155}
\end{equation}
In order to establish the exact numerical prefactor in front of the last term
in (\ref{GZ155}) we also took into account the $T$-dependent linear in time
contribution to $A(t)$ \cite{GZ99} omitted in eq. (\ref{Acorr}). We note that
the result (\ref{GZ155}) agrees (up to a numerical prefactor of order one
in front of the term $\propto T\tau_H^2$) with the perturbative result \cite{AAG2}
obtained in the same limit $T \tau_H \gg 1$. However, as it was already
discussed above, no information about the dephasing time at low $T$ can be
extracted from (\ref{GZ155}).

Finally, let us derive the expression for the weak localization correction in
the limit of high temperatures and weak magnetic fields 
$\tau_H \gg \tau_{\varphi}$. In this limit the phase relaxation is determined 
by the term $f_{cl} \propto Tt^{3/2}$ in (\ref{fthermal}). Keeping this term in the 
exponent, expanding the exponent in $f(t)-f_{cl}(t)$ to the first order and
making use of a short time expansion of $A(t)$ we get 
\begin{equation}
\delta\sigma_1 \simeq  -\left(\frac{2e^4\sigma_1D}{9\pi^4T}\right)^{1/3}\Gamma (1/3)
+\frac{\zeta (1/2)}{(2\pi )^{3/2}}
\frac{e^2\sqrt{D}}{\sqrt{T}}.
\label{chVA3}
\end{equation}  
Here $\Gamma (x)$ is the Euler's gamma-function. The first term in the right-hand
side of this expression corresponds to the classical result \cite{AAK}, while
the second term represents the first correction to this result. 
With decreasing temperature this correction grows faster than the absolute value
of the first term. At the same time in the high temperature limit the second
term in (\ref{chVA3}) always remains much smaller than 
the first one and the higher order terms should also be taken into 
account already at $T>T_q$.

Note that the result (\ref{chVA3}) -- as well as eqs. (\ref{GZ156}),
(\ref{GZ155}) -- does not depend
on the zero temperature value of $\tau_{\varphi}$, this value drops out 
due to the same cancellation
of the linear in time $T$-independent terms from the exponent and the pre-exponent.
Thus also the first order high temperature expansion of the weak 
localization correction
to the conductance cannot provide correct information about the 
electron dephasing time at low temperatures.

In conclusion, we established an explicit relation between
perturbative and nonperturbative results in the problem of
quantum dephasing in disordered conductors. The dephasing time
remains finite down to $T=0$ due to the electron-electron interactions.


\begin{thebibliography}{99}

\bibitem{Webb} P. Mohanty, E.M.Q. Jariwala, and R.A. Webb, Phys. Rev. Lett.
{\bf 78}, 3366 (1997); Fortsch. Phys. {\bf 46}, 779 (1998).
\bibitem{AAK} B.L. Altshuler, A.G. Aronov, and D.E. Khmelnitskii,
J. Phys. C {\bf 15}, 7367 (1982).
\bibitem{AAK1} B.L. Altshuler and A.G. Aronov, in {\it Electron-Electron
Interactions in Disordered Systems}, edited by A.L. Efros and M. Pollak
(North-Holland, Amsterdam, 1985), p.1.
\bibitem{Mohanty} P. Mohanty, this volume and further Refs. therein.
\bibitem{GZ} D.S. Golubev and A.D. Zaikin, Phys. Rev. Lett. 
{\bf 81}, 1074 (1998); Phys. Rev. B {\bf 59}, 9195 (1999). 
\bibitem{GZ98} D.S. Golubev and A.D. Zaikin, Physica B {\bf 225}, 164 (1998).
\bibitem{CL} A.O. Caldeira and A.J. Leggett, Physica A {\bf 121}, 
587 (1983); {\bf 130}, 374 (1985).
\bibitem{LM} D. Loss and K. Mullen, Phys. Rev. B {\bf 43}, 13252 (1991).
\bibitem{Weiss} U. Weiss, {\it Quantum Dissipative Systems}
(World Scientific, Singapore, second enlarged edition, 1999).
\bibitem{AGA} B.L. Altshuler, M.E. Gershenzon, and I.L. Aleiner, Physica E
{\bf 3}, 58 (1998).
\bibitem{AAG2} I.L. Aleiner, B.L. Altshuler, and M.E. Gershenzon, Phys. 
Rev. Lett. {\bf 82}, 3190 (1999); cond-mat/9808053.
\bibitem{GZr} D.S. Golubev and A.D. Zaikin, Phys. Rev. Lett. {\bf 82}, 3191 (1999);
cond-mat/9811185.
\bibitem{Naz} Yu. V. Nazarov, Sov. Phys. JETP {\bf 68}, 561 (1989);
M.H. Devoret {\it et al.} Phys. Rev. Lett. {\bf 64}, 
1824 (1990); S.M. Girvin {\it et al.}, {\it ibid.} {\bf 64}, 1565 (1990).
\bibitem{SZ} G. Sch\"on and A.D. Zaikin, Phys. Rep. {\bf 198}, 237 (1990).
\bibitem{many} L.S.Kuzmin {\it et al.},
Phys. Rev. Lett. {\bf 67}, 1161 (1991); D.B. Haviland {\it et al.},
Europhys. Lett. {\bf 16}, 103 (1991); A.N. Cleland, J.M. Schmidt, and 
J. Clarke, Phys. Rev. B. {\bf 45}, 2950 (1992);  P. Joyez, D. Esteve, and 
M.H. Devoret, Phys. Rev. Lett. {\bf 80} 1956 (1998).
\bibitem{GZPE} D.S. Golubev and A.D. Zaikin, to appear 
in: {\it Quantum Physics at
Mesoscopic Scale} eds. D.C. Glattli, M. Sanquer and J. Tran Thanh Van 
(Frontieres, 1999); cond-mat/9907497.
\bibitem{FA} H. Fukuyama and E. Abrahams, Phys. Rev. B {\bf 27}, 5976 (1983).
\bibitem{GZ99} D.S. Golubev and A.D. Zaikin, cond-mat/9907494.
\bibitem{CS} S. Chakravarty and A. Schmid, Phys. Rep. {\bf 140}, 193 (1986).
\end{thebibliography}
\end{document}